\documentclass[aps,prb,twocolumn,groupedaddress]{revtex4}
\usepackage{amsmath,amssymb}
\usepackage{graphics}
\usepackage{bm}
\usepackage{epsfig}
\usepackage{subfig}
\usepackage{amsfonts}

\begin{document}

\title{Induced p-wave Superfluidity in Imbalanced Fermi Gases in a Synthetic Gauge Field}

\author{Heron Caldas}\email{hcaldas@ufsj.edu.br}
\affiliation{Departamento de Ci\^{e}ncias Naturais, Universidade Federal de
  S\~ao Jo\~ao Del Rei, 36301-160, S\~ao Jo\~ao Del Rei, MG, Brazil}

\author{Mucio Continentino} \email{mucio@cbpf.br} \affiliation{Centro Brasileiro de Pesquisas Fisicas, Rua Dr. Xavier Sigaud, 150, Urca 
22290-180, Rio de Janeiro, RJ, Brazil}

\date{\today}

\begin{abstract}

We study pairing formation and the appearance of induced spin-triplet p-wave superfluidity in dilute three-dimensional imbalanced Fermi gases in the presence of a uniform non-Abelian gauge field. This gauge field generates a synthetic Rashba-type spin-orbit interaction which has remarkable consequences in the induced p-wave pairing gaps. Without the synthetic gauge field, the p-wave pairing occurs in one of the components due to the induced (second-order) interaction via an exchange of density fluctuations in the other component. We show that this p-wave superfluid gap induced by density fluctuations is greatly enhanced due to the Rashba-type spin-orbit coupling.

\end{abstract}


\maketitle

\section{Introduction}
\label{int}

The tremendous improvement of the techniques of dealing with ultra cold atoms in the last few years, has paved the way for studying many-body quantum phenomena in an unprecedented manner~\cite{Exp1,Exp2,Exp3,Exp4}. The possibility of observation of exotic states of matter, which may have analogies with condensed matter, quark matter and neutron star physics, such as topological phase transitions (TPT), Majorana fermions and color superconductivity, has greatly motivated the investigation of pairing and condensation of ultra cold Fermi systems under the influence of external electric and magnetic fields~\cite{Bloch,Giorgini}. While the laser field allows the spin-orbit coupling, the Zeeman magnetic field leads to an imbalance between the spin up and spin down chemical potential of the two fermionic species. TPT emerge only in the presence of these two fields~\cite{ReviewAlicea}.

Motivated by recent experimental realization of synthetic spin-orbit coupling for ultracold atoms~\cite{sint1,sint2,sint3}, we study the ground state of dilute (spin $1/2$) Fermi gases. We investigate the manifestation of two possible induced p-wave pairing gaps, a ``direct'' one. induced by a Rashba-type spin-orbit coupling (RTSOC) generated by a synthetic gauge field~\cite{Shenoy1,Nandini}, and that induced by density fluctuations.

From the cold atom side, pairing formation in non-conventional systems (e.g. the one formed by a two-species imbalanced configuration) is of great interest in the investigation of mixtures of alkali atoms as, for example, Lithium-Potassium mixtures~\cite{Petrov}. A multi-pairing system turns out to be very interesting when compared to the usual configuration, since the ground state now results from a competition not only of Fermi surface, chemical potential and mass mismatches, but also from the several pairing gaps that can simultaneously be present~\cite{Bedaque,Torma}.

However, these recent investigations in imbalanced (non-conventional) systems without RTSOC did not consider interactions between the same component. In other words, only s-wave (inter-species) pairing gaps have been taken into account. In this paper we consider pairing between the same species from the point of view of induced interactions that can emerge in the two situations we mentioned earlier.

In an imbalanced configuration, standard (BCS) s-wave pairing is energetically unfavorable to occur~\cite{Caldas1,Caldas2}. In this adverse scenario, other kinds of pairing are expected to manifest~\cite{K,L}. Intra-species p-wave pairing gaps induced by density fluctuations are our main interest in this work, since those gaps survive in the limit of vanishing RTSOC and chemical potential imbalance. Previous studies (without RTSOC) found that while the p-wave energy gain is parametrically smaller in weak coupling, in asymptotic regions of imbalance and coupling the gaps are exponentially suppressed~\cite{Aurel}. We find that this is modified in the presence of a RTSOC. We show that even in weak coupling, the p-wave pairing gap induced by density fluctuations can be greatly enhanced, due to RTSOC.

The possibility of the Fulde-Ferrel-Larkin-Ovchinnikov (FFLO) state with modulated order parameter~\cite{FFLO} is ignored in this work. We consider only pairing between atoms with equal and opposite momenta.

This paper is organized as follows. In Sec.~I, after introducing the model, we obtain a diagonal (in the helicity basis) Hamiltonian, which contais the free and the RTSOC terms. In Sec.~II we calculate the p-wave pairing gaps in the presence of a RTSOC and by density fluctuations as a function of this RTSOC, and investigate its effects on the p-wave gaps. We conclude in Sec.~III.

\section{The Model}
\label{sec2}

\begin{figure}[th]
\centering{\includegraphics[scale=0.38,angle=90]{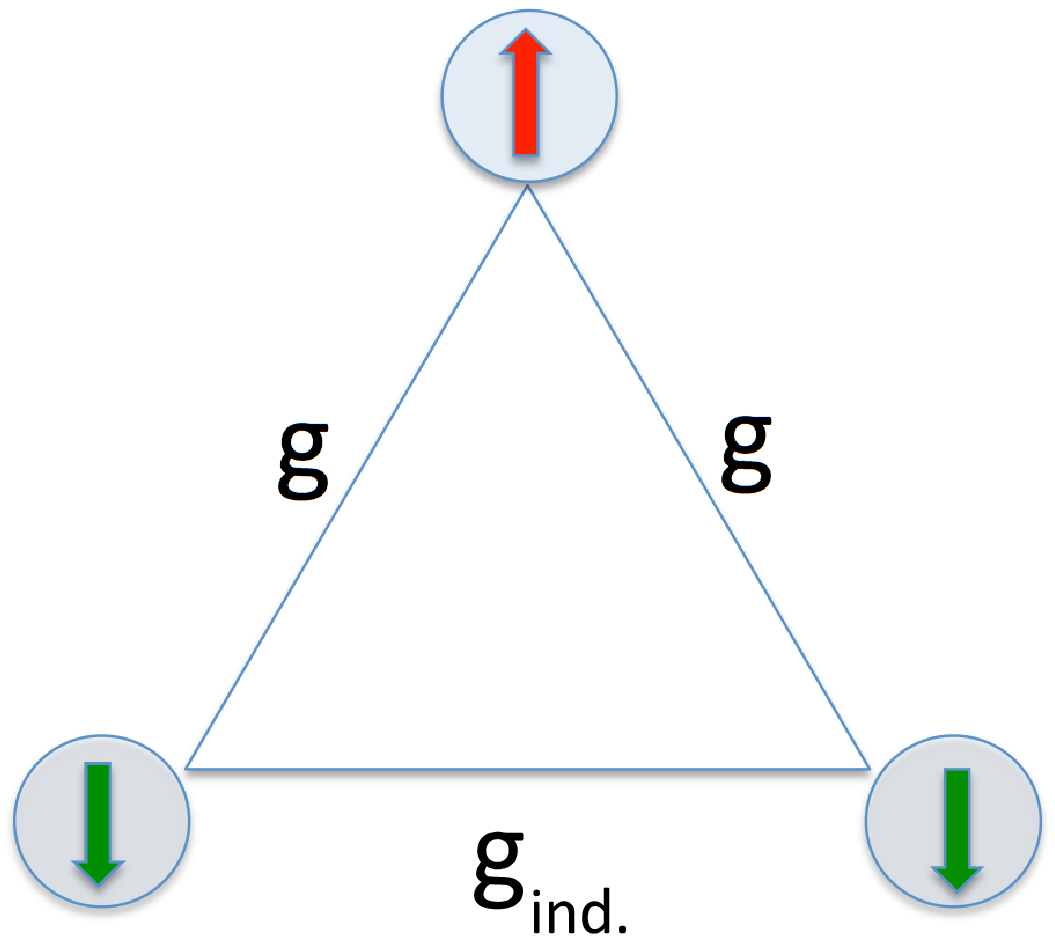}}\caption{(Color online) Spin-up atoms form s-wave pairing with spin-down ones, while induced p-wave pairing between atoms of same species emerges due to the presence of the other species.}
\label{fig1}
\end{figure}

The system we are investigating is an imbalanced Fermi gas which is illustrated by Fig.~\ref{fig1}. An intra-species pairing gap between atoms of same species (both of spin down) will naturally emerge, due to interactions induced by density fluctuations, even in the absence of a Rashba-type spin-orbit coupling. This pairing gap is of p-wave, since s-wave induced interactions is forbidden by Pauli exclusion. As we will see below, additional p-wave pairing gaps will manifest, induced by a RTSOC. 

To model the (s-wave) interaction between the spin-up and spin-down atoms, we consider a uniform 3D polarized Fermi gas with a RTSOC, described by the Hamiltonian:

\begin{eqnarray}
H=H_{0}+H_{SO}+H_{int}, \label{1}
\end{eqnarray}
where $H_{0}$ is the kinetic term, $H_{SO}$ is a Rashba-type spin-orbit interaction, generated by a synthetic gauge field, and $H_{int}$ is the term with a short-range s-wave interaction between the two fermionic species.

\begin{eqnarray}
&H_{0}&=\sum_{\textbf{k},\sigma} \xi_{\textbf{k},\sigma} c_{\textbf{k},\sigma}^{\dag} c_{\textbf{k},\sigma}, \nonumber\\
&H_{SO}&=\sum_{\textbf{k}} \lambda k \left(e^{-i \varphi_{\textbf{k}}} c_{\textbf{k},\uparrow}^{\dag} c_{\textbf{k},\downarrow} + 
h.c. \right), \nonumber\\
&H_{int}&= 
\sum_{\textbf{k},\textbf{k}'} g(\textbf{k},\textbf{k}^{\prime}) c_{\textbf{k},\uparrow}^{\dag}
c_{-\textbf{k},\downarrow}^{\dag} c_{-\textbf{k}',\downarrow}
c_{\textbf{k}',\uparrow}, 
\label{2}
\end{eqnarray}
with $\xi_{\textbf{k},\sigma} =\epsilon_k - \mu_{\sigma} =\hbar k^{2}/(2m)-\mu_{\sigma}$, where $\mu_{\sigma=\uparrow, \downarrow} = \mu \pm h$ is the chemical potential of the $\sigma=\{ \uparrow, \downarrow \}$ species. Here $\mu$ is the chemical potential of the balanced system and $h$ is an effective Zeeman field. We consider the imbalanced configuration in which $k_F^{\uparrow}>k_F^{\downarrow}$, where $k_F^{\uparrow,\downarrow}=\sqrt{2m\mu_{\uparrow,\downarrow}}$ is the Fermi momentum of the $\uparrow,\downarrow$ species, respectively. $c_{\textbf{k},\sigma}^{\dag}(c_{\textbf{k},\sigma})$ denotes the creation(annihilation) operators for a fermion with momentum
$\textbf{k}$ and spin $\sigma$, $\lambda$ is the strength of gauge field configuration, with $\varphi_{\textbf{k}} = \arg(k_{x}+ i k_{y})$ and $g(\textbf{k},\textbf{k}^{\prime})$ is the interaction potential (for convenience, we set $\hbar=k_B=1$). Notice that in Eq.~(\ref{2}) $k=\sqrt{k_x^2+k_y^2+k_z^2}$. This corresponds to the most symmetric synthetic gauge field configuration~\cite{Shenoy1,Lin}. In two dimensions this RTSOC interaction, generated by a synthetic gauge field, reduces to the usual Rashba spin-orbit coupling term.

The presence of the interaction term in $H$ is responsible for the appearance of non-trivial spatial correlations between the fermions of spin $\sigma$ and $\sigma'$, which are proportional to a gap parameter $\Delta_{\sigma \sigma'}(\textbf{k}) \sim \langle c_{\textbf{k},\sigma} c_{-\textbf{k},\sigma'}  \rangle$, where $\sigma,\sigma'=\{ \uparrow, \downarrow \}$. The Pauli exclusion principle imposes that the gap parameter has to obey $\Delta_{\sigma \sigma'}(\textbf{k})=-\Delta_{\sigma' \sigma}(-\textbf{k})$. There are four possible spin configurations for the two spin $1/2$ fermions considered here: one singlet and three triplets. The gap parameters may be written as $\Delta_{\sigma \sigma'}(\textbf{k}) \sim \sum_{l,m} \Delta_{l,m} Y_{l,m}$, where $Y_{l,m}$ are the spherical harmonics. The p-wave triplet refers to angular momentum $ l=1$, for which the magnetic moments are $m=0,\pm1$. The spherical harmonics $Y_{l,m}$ with $l=1$ can be expressed as a linear function of the vector $\textbf{k}$ as~\cite{Book}, $Y_{1,1}(\textbf{k}) \sim k_x+ik_y$, $Y_{1,-1}(\textbf{k}) \sim k_x-ik_y$, and $Y_{1,0}(\textbf{k}) \sim k_z$, where $Y_{1,\pm 1}(\textbf{k})$ represent a state with two nodal points, while $Y_{1,0}(\textbf{k})$ a state with a nodal plane. The states with order parameter $\Delta_{\sigma \sigma'}(\textbf{k}) \sim Y_{1,\pm 1}(\textbf{k})$ break time-reversal symmetry. As we will see below, the two induced p-wave triplet pairing gaps we find here are of this type.

It is instructive to diagonalize first $H_{0}+H_{SO}$, which gives

\begin{eqnarray}
H_{0}+H_{SO} &=& \sum_{\textbf{k}, j=\pm} \xi_{\textbf{k},j}
a_{\textbf{k},j}^{\dag} a_{\textbf{k},j}.
\label{4}
\end{eqnarray}
In Eq.~(\ref{4}) $a^{\dag}_{\textbf{k},\pm}(a_{\textbf{k},\pm})$ is the creation(annihilation) operator for the state with helicity $(\pm)$, $\xi_{\textbf{k},\pm}=\xi_{\textbf{k}} \pm H_k$, where $\xi_{\textbf{k}}=k^{2}/(2m)-\mu$, and $H_k \equiv \sqrt{h^2+\lambda^2 k^2}$.

In the next section we obtain the p-wave gap induced by a RTSOC, and calculate that induced by density fluctuations. Finally we discuss the relative importance of these contributions in a unique expression.

\section{The triplet pairing gaps}
\label{sec2}

\subsection{P-wave triplet pairing induced by a RTSOC}

Since we are mainly interested to investigate induced (by both a RTSOC and density fluctuations) p-wave triplet pairings, we take a constant (i.e., $k$-independent) interaction strength $g<0$, which is equivalent to consider only spin-singlet pairing. We treat the $H_{int}$ term using a BCS decoupling, and then write it in the basis that diagonalizes $H_{0}+H_{SO}$~\cite{Alicea},

\begin{eqnarray}
\label{HMF}
H_{int} = &-&\frac{\mid \Delta \mid^{2}}{g} + \sum_{\textbf{k}} \Delta_{+ -}(k) a_{\textbf{k},+}^{\dag}
a_{-\textbf{k},-}^{\dag}\\
\nonumber  
&+& \sum_{\textbf{k},j=\pm} \Delta_{j j}(k) a_{\textbf{k},j}^{\dag}
a_{-\textbf{k},j}^{\dag}  + h.c.
\end{eqnarray}

To regulate the divergence associated with the contact interaction term in $H_{int}$, we use

\begin{equation}
\label{renorm}
\frac{1}{g}=\frac{m}{4\pi a_s} -  \int \frac{d^3 k}{(2 \pi)^3} \frac{m}{k^2}.
\end{equation}
Besides the regularization of the ultraviolet divergence present in the gap equation, this equation relates the strength $g$ of the contact interaction with the three dimensional scattering length $a_s$, which is more physically relevant, since it permits to make contact with current experiments.

In Eq.~(\ref{HMF}) we have defined $\Delta_{+ -}(k)=\Delta_s(k)$, and~\cite{Alicea}

\begin{eqnarray}
\label{dominant}
\Delta_{+ +}(k)=\frac{(k_x+ik_y)}{k_{\perp}} \Delta_p\\
\nonumber  
\Delta_{- -}(k)=\frac{(k_x-ik_y)}{k_{\perp}} \Delta_p,
\end{eqnarray}
where $k_{\perp}=\sqrt{k_x^2+k_y^2}$, and

\begin{eqnarray}
\left(
  \begin{array}{c}
    \Delta_s(k)\\
    \Delta_p(k)\\
  \end{array}
\right)=\frac{1}{2 \sqrt{\lambda^2 k^2+h^2}}
\left(
  \begin{array}{c}
    2h\\
    -\lambda k\\
  \end{array}
\right)
\Delta.
\label{gaps}
\end{eqnarray}
$\Delta= -g \sum_{\textbf{k}} <c_{-\textbf{k},\downarrow} c_{\textbf{k},\uparrow}>$ is the s-wave energy gap. Notice that $\Delta_s(k)^2+4\Delta_p(k)^2= \Delta^2$. This shows that in an imbalanced ($h \neq 0$) Fermi system with RTSOC, the original s-wave order parameter ($\Delta$) has contributions from pairing with both the same and different helicity states~\cite{JSTAT}.

The pairing gaps $\Delta_{-  -}(k)$ and $\Delta_{+ +}(k)$ are of the p-wave triplet type ($\sim Y_{1,\pm 1}(\textbf{k})$) as we mentioned earlier. However, they vanish in the limit $\lambda \to 0$. This is the reason we have named them as ``induced by RTSOC'' p-wave pairing gaps.

The excitation spectra read $E_{\textbf{k},\pm}=\sqrt{\xi_{\textbf{k}}^{2} + \mid \Delta \mid^{2} + H_k^2  \pm 2 E_0}$, where $E_0= \sqrt{h^2(\xi_{\textbf{k}}^{2} + \mid \Delta \mid^{2}) + \lambda^2 k^2\xi_{\textbf{k}}^{2} }$. 
The vanishing of the excitation energies (i.e. $E_{\textbf{k},\pm}=0$) implies the equation

\begin{eqnarray}
(E_k^2-H_k^2)^2 + (2\lambda k \Delta)^2=0,
\label{ve}
\end{eqnarray}
where $E_k=\sqrt{\xi_{\textbf{k}}^2+ \Delta^2}$. Notice that the above equation will be satisfied only for $k=0$ and $E_{k=0}=H_{k=0}$. This gives an equation for a critical field $h_c=\sqrt{\mu^2+ \Delta^2}$ at which a quantum phase transition to a topological superfluid state occurs~\cite{LHe,Zhang}. For a fixed value of $\lambda$ and in the limit $h \gg h_c$ it is found that the pairing gap which enters in Eq.~(\ref{gaps}) behaves as~$\Delta \sim C/h^2$ (where $C$ is a constant depending on $\lambda$) both in 2D~\cite{LHe} and 3D~\cite{Zhang}. Thus, in a highly imbalanced system, the ``direct'' triplet p-wave pairing gaps decrease as~$\Delta_p \sim C/h^3$. Therefore, in such an imbalanced configuration, when the imbalance tends to lead the system to the normal state, the direct triplet p-wave pairing gaps are vanishingly small. This is natural to expect, since the spin-triplet p-wave pairing induced by RTSOC occurs due to the direct short-range s-wave inter-component interaction~\cite{Gor}.
% because of nonconservation of spin: the pair wavefunction is a mixture of spin-singlet and spin-triplet channels

As we will see next, there is a second-order interaction, that also generates a p-wave pairing gap, which is that induced by density fluctuations. This investigation has been carried out previously without RTSOC in, for instance, Refs.~\cite{Aurel} and~\cite{Dan}. It was found that the p-wave pairing gaps induced by density fluctuations are exponentially suppressed by both the strength of the coupling constant and asymmetry between the up and down species. We show that they are significantly enhanced when the effects of RTSOC are taken into account. More importantly, we discuss the limits the p-wave pairing gaps induced by density fluctuations overcome that from first-order contributions.

\subsection{P-wave triplet pairing induced by density fluctuations}

\begin{figure}
\includegraphics[scale=0.38,angle=90]{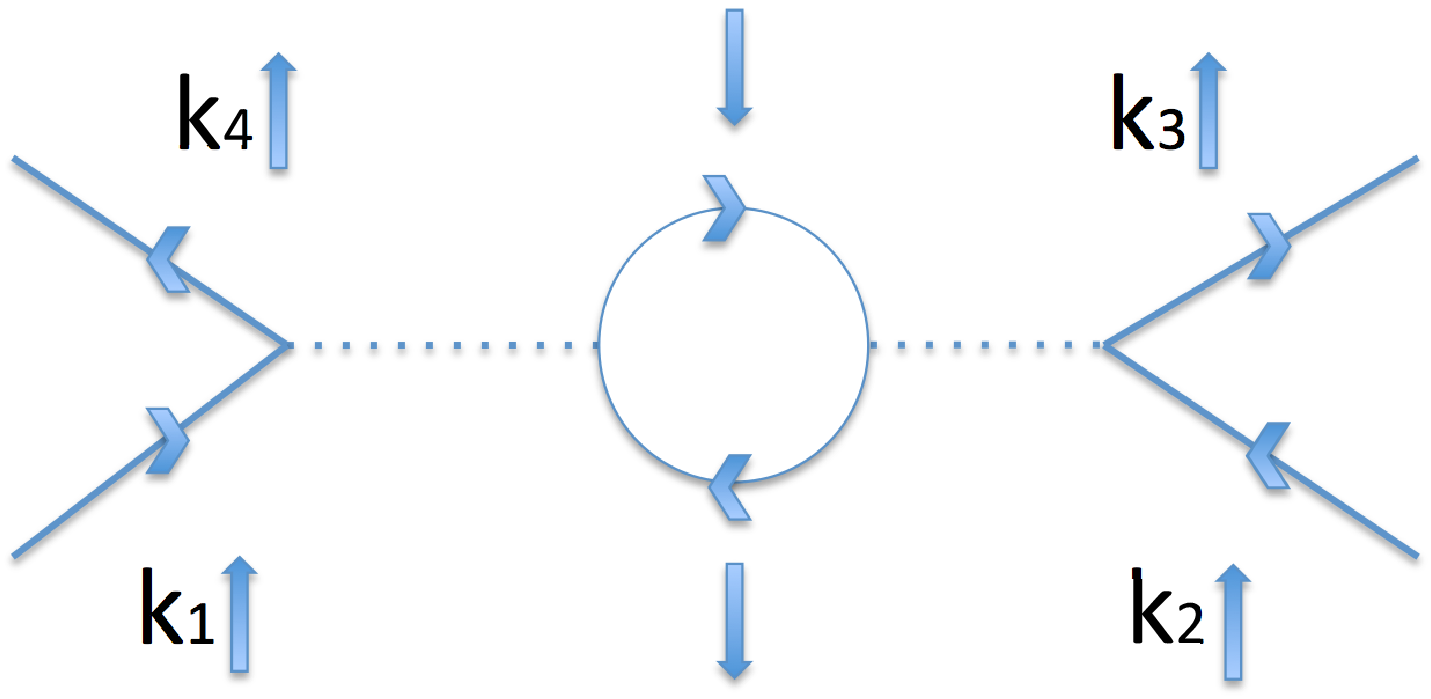}
\caption{Induced interaction for back-to-back scattering in the original spin representation.} 
\label{fig2}
\end{figure}

We start the calculation of the induced interaction for the majority species, i.e., for the atoms with $+$ helicity. The induced interaction was obtained originally by Gorkov and Melik-Barkhudarov (GMB) in the BCS limit by second-order perturbation theory~\cite{GMB61}. For a (back-to-back) scattering process depicted in Fig.~\ref{fig2}, with $p_1+p_2\rightarrow p_3+p_4$, the induced interaction to lowest order in the s-wave chanel is given by~\cite{Pethick00},

\begin{equation}
\label{IndInt}
U_{\mathrm{ind}}( p_1, p_4)= -g^2\, \chi_{ph}(p_1-p_4),
\end{equation}
where $p_{i}=({\bf k}_{i}, \omega_{l_i})$ is a vector in the space of wave-vector ${\bf k}$ and fermion Matsubara frequency $\omega_{l}=(2l+1)\pi/(\beta)$. The polarization function $\chi_{ph}(p')$ is given by

\begin{eqnarray}
\chi_{ph}(p')&=& \frac{1}{ \beta {\rm V}}\sum_p
\mathcal{G}_{0 b}(p)\mathcal{G}_{0 a}(p+p')\\
\nonumber
&=&\int \frac{\rm{d}^3 {\bf k}}{(2\pi)^3} \frac{f_{{\bf k},-} -f_{{\bf k}+{\bf q},-}}{ i
\Omega_l+\xi_{{\bf k},-} -\xi_{{\bf k}+{\bf q},-}},
\label{chi0}
\end{eqnarray}
where $p'=(\bf{q}, \Omega_{l})$, $\Omega_{l}=2l\pi/\beta$ is the Matsubara frequency of a boson. The Matsubara Green's function of a non-interacting Fermi gas is given by $\mathcal{G}_{0 -}(p)=1/(i\omega_l-\xi_{\textbf{k},-}) $. Notice that this polarization function is being calculated in the basis of the helicity states. Making the integrations above assuming that $h\ll \lambda k_F$, we get

\begin{eqnarray}
\label{chiph-9}
\chi(|\vec q|) = - N_{F,\lambda}^{-} L(x^-) - N_{\lambda} [ L(z) + F(x^-)-F(z) ],
\end{eqnarray}
where $N_{F,\lambda}^{-}=\frac{m k_{F,\lambda}^{-}}{2 \pi^2}$ is the density of states at the ``deformed'' Fermi surface of the $-$ species, $N_{\lambda}=\frac{m k_{\lambda}}{2 \pi^2}$ , $x^- = \frac{q}{2 k_{F,\lambda}^{-}}$, $z=\frac{q}{2 k_{\lambda}}$, $k_{F,\lambda}^{-}=k_{F}^{-}-k_{\lambda}$, with $k_{\lambda} \equiv m \lambda$, and  $q=|\vec q|$. $L(x)$ is the static Lindhard function,

\begin{eqnarray}
\label{chiph9-2}
L(x)= \frac{1}{2} -\frac{1}{4x}(1-x^2) \ln \left|\frac{1-x}{1+x} \right|,
\end{eqnarray}
and we have also defined

\begin{eqnarray}
\label{FL}
F(x)= - \frac{1}{x} \ln \left|\frac{1-x}{1+x} \right| + \ln  \left|1-\frac{1}{x^2} \right|.
\end{eqnarray}
The Fermi momenta of the (free) $+-$ atoms are found by $\xi_{k_F,\pm}=0$, which read~\cite{Alicea}

\begin{equation}
\label{kf1}
k_{F,\pm}= \sqrt{2m(\mu+m \lambda^2) \pm 2m \sqrt{h^2+ m \lambda^2 (m \lambda^2+ 2\mu)}}.
\end{equation}
Expressing the coupling $g$ in terms of the s-wave scattering length to lowest order, 

\begin{equation}
U_{\mathrm{ind}}= \left(\frac{4 \pi a_s}{m} \right)^2 [ N_{F,\lambda}^{-} ~ L(x^-)+ N_{\lambda} (L(z) + F(x^-)-F(z))].
\end{equation}

In the scattering process conservation of momentum implies $\vec{k}_1+\vec{k}_2=\vec{k}_3+\vec{k}_4$ which is set to zero. $q$ is equal to the magnitude of $\vec{k}_1+\vec{k}_3=\vec{k}_1-\vec{k}_4$, then 
$q=\sqrt{(\vec{k}_1+\vec{k}_3).(\vec{k}_1+\vec{k}_3)}=\sqrt{\vec{k}_1^2+\vec{k}_3^2+ 2\vec{k}_1. \vec{k}_3}=\sqrt{\vec{k}_1^2+
\vec{k}_3^2+ 2|\vec{k}_1|| \vec{k}_3|\cos\theta}$. Since both particles are at Fermi surface the atoms with $+$ helicity, $|\vec{k}_1|=|\vec{k}_3|=k_F^{+}$, thus, 
$q=k_F^{+}\sqrt{2(1+\cos \theta)}$. Then we have $x^-=\frac{q}{2k_{F,\lambda}^{-}}=\frac{k_F^{+}}{k_{F,\lambda}^{-}} \frac{\sqrt{2(1+\cos \theta)}}{2} = y \frac{\sqrt{2(1+\cos \theta)}}{2}$, where $y=\frac{k_F^{+}}{k_{F,\lambda}^{-}}$, and $z=\frac{q}{2k_{\lambda}}=\frac{k_F^{+}}{k_{\lambda}} \frac{\sqrt{2(1+\cos \theta)}}{2} = \eta \frac{\sqrt{2(1+\cos \theta)}}{2}$, with $\eta=\frac{k_F^{+}}{k_{\lambda}}$. 

Taking the projection onto the Legendre polynomial $P_{l=1} (\cos (\theta))$~\cite{Aurel,Dan,Leo}

\begin{equation}
U_{P} = \frac{1}{2} \int_{0}^{ \pi} \cos(\theta) \sin(\theta) d  \theta ~U_{ind}(x),
\end{equation}
we find

\begin{equation}
U_{P}= -\left(\frac{4 \pi a_s}{m} \right)^2 G_1(h,\alpha),
\end{equation} 
where $G_1(h,\alpha) = N_{F,\lambda}^{-} L_1(y) + N_{\lambda} [L_1(\eta)+F_1(y)-F_1(\eta)]$, with the following definitions

\begin{eqnarray}
\label{L1}
&&L_1(x)= \\
\nonumber
&&\frac{5 x^2-2}{15 x^4} \ln \left| 1-x^2 \right|  -\frac{x^2+5}{30x} \ln \left|\frac{1-x}{1+x} \right| -\frac{x^2+2}{15 x^2},
\end{eqnarray}
and 

\begin{eqnarray}
\label{F1}
&&F_1(x)= \\
\nonumber
&&  2 \ln \left|\frac{1-x}{1+x} \right| -x \ln \left| 1-\frac{1}{x^2} \right| -\frac{1}{x} \ln \left|1-x^2 \right|.
\end{eqnarray}
It is worth to mention that induced interactions have been used to obtain the transition temperature (or tricritical point) beyond mean-field both in three~\cite{Pethick00,Baranov2,Yu10}, and two dimensions~\cite{Baranov,Caldas1,Caldas2}. We can express $y$ and $\eta$ in terms of the modified by RTSOC Fermi vectors $k_{F,\pm}$,

\begin{eqnarray}
\label{y}
y=\frac{\sqrt{1+2\alpha^2 + \sqrt{\bar h^2+2\alpha^2(2\alpha^2+2)}}}{\sqrt{1+2\alpha^2 - \sqrt{\bar h^2+2\alpha^2(2\alpha^2+2)}}-\alpha},
\end{eqnarray}

\begin{eqnarray}
\label{eta}
\eta=\frac{1}{\alpha} \sqrt{1+2\alpha^2 + \sqrt{\bar h^2+2\alpha^2(2\alpha^2+2)}}.
\end{eqnarray}
Here we have defined $\alpha=\lambda/v_F$, where $v_F$ is the Fermi velocity, and $\bar h= h/\mu$. The Fermi vector $k_{F,-}$ and $y$ are both real for $h<\mu$, as is the case without RTSOC. An analogy with the mean-field analysis~\cite{Aurel} leads to the (second-order) p-wave superfluid triplet intra-species pairing amplitude,

\begin{eqnarray}
\label{pairamp}
\tilde{\Delta}_{+ +} \sim {E_F^{+}}  {\rm exp}  \left[- \frac{ \pi^2 }{4 ~a_s^2~  k_F^{+} k_{F,\lambda}^{-}  ~  f_1(h,\alpha)}\right],
\end{eqnarray} 
where ${E_F^{+}}$ is the Fermi energy of the fermions with $+$ helicity, $f_1(h,\alpha) = L_1(y) + \frac{N_{\lambda}}{N_{F,\lambda}^{-} } [L_1(\eta)+F_1(y)-F_1(\eta)]$. To leading order in $\alpha$, we can approximate $k_F^{+} k_{F,\lambda}^{-}  ~  f_1(h,\alpha) \sim
(  k_F^{+} k_{F}^{-} - k_F^{+} k_{\lambda} )    L_1(y) = k_F^2 ~ g_3(h,\alpha)  L_1(y) $, where we have defined $g_3(h,\alpha) = g_1(h,\alpha)-g_2(h,\alpha)$, with $g_1(h,\alpha)=\sqrt{1+4\alpha^4 -\bar h^2}$ and $g_2(h,\alpha)=\alpha \sqrt{1+2\alpha^2 + \sqrt{\bar h^2+2\alpha^2(2\alpha^2+2)}}$. Finally we obtain

\begin{eqnarray}
\label{gapmais}
\tilde{\Delta}_{+ +} \sim E_F^{+} {\rm exp}  \left[- \left( \frac{ \pi }{2 ~  k_F  ~ a_s} \right)^2 \frac{1}{g_3(h,\alpha)   L_1(y)}\right].
\end{eqnarray}

Since the lower branch is being emptied by increasing te RTSOC strength, the induced pairing between atoms of the {\it negative} helicity is strongly suppressed. Its given by

\begin{eqnarray}
\label{gapmenos}
\tilde{\Delta}_{- -} \sim E_F^{-} {\rm exp}  \left[- \left( \frac{ \pi }{2 ~  k_F  ~ a_s} \right)^2 \frac{1}{g_3(h,\alpha)   L_1(y_2)}\right],
\end{eqnarray}
where

\begin{eqnarray}
\label{y2}
y_2=\frac{\sqrt{1+2\alpha^2 - \sqrt{\bar h^2+2\alpha^2(2\alpha^2+2)}}}{\sqrt{1+2\alpha^2 + \sqrt{\bar h^2+2\alpha^2(2\alpha^2+2)}}+\alpha}.
\end{eqnarray}
The results of Ref.~\cite{Aurel} are readily obtained in the limit $\alpha=0$, since $k_F^2 g_3(h,\alpha=0)= k_F^{\uparrow} k_{F}^{\downarrow}$, $y(\alpha=0)=k_F^{\uparrow} / k_{F}^{\downarrow}$, and $y_2(\alpha=0)=y(\alpha=0)^{-1}=k_{F}^{\downarrow} / k_F^{\uparrow}$.

\begin{figure}[th]
\centering{\includegraphics[scale=0.41,angle=-90]{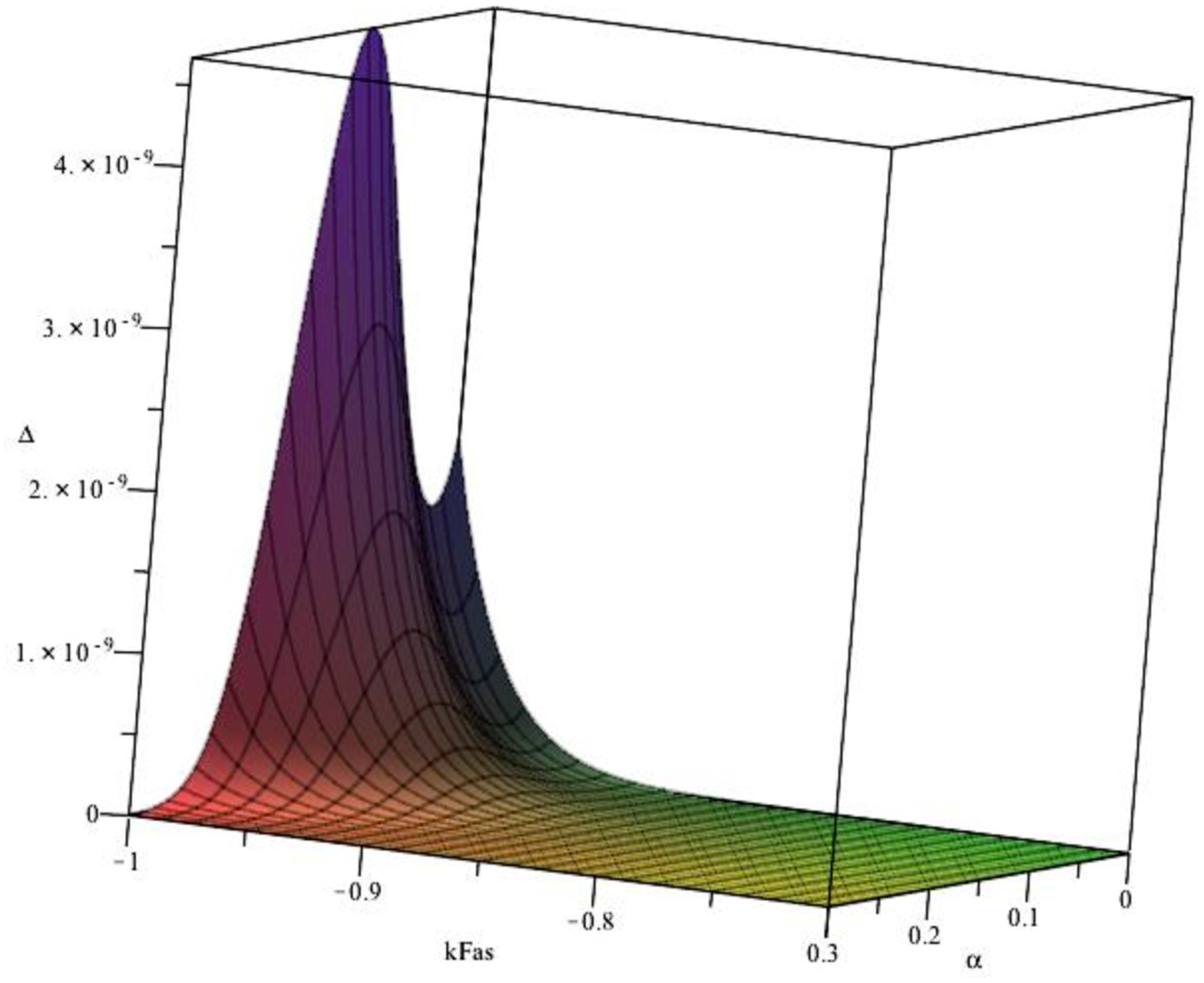}}\caption{(Color online) Behavior of the upper branch helicity induced p-wave pairing gap over the Fermi energy, $\Delta \equiv \tilde{\Delta}^{+ +}/E_F^{+}$, as a function of the non-dimensional parameters $k_F  a_s$ and $\alpha$.}
\label{Delta}
\end{figure}

In Fig.~\ref{Delta} we show the behavior of $\tilde{\Delta}_{+ +}/E_F^{+}$ as a function of $k_F  a_s$ and $\alpha$ for $\bar h=0.15$. Notice that $\tilde{\Delta}_{+ +}/E_F^{+}$ increases with $k_F  a_s$ in the BCS range considered and has a maximum for $\alpha \sim 0.14$ for all values of $k_F  a_s$. This should be compared with the maximum value of $\Delta^{\uparrow \uparrow}/E_F^{\uparrow}$ obtained for $\alpha=0$ in Ref.~\cite{Aurel}, for the same value of $k_F a_s=-0.8$, $\Delta^{\uparrow \uparrow}/E_F^{\uparrow} \sim 10^{-16}$, which is $10^{2}$ times lower. Notice that the absolute value of the $\tilde{\Delta}_{+ +}$ can be increased by increasing $E_F^{+}$, i.e., considering systems with higher densities and/or lighter particles.

With the induced interaction between intra-species atoms, Eq.~(\ref{IndInt}), we can make a BCS-type calculation and obtain the second-order triplet pairing gaps (not only their amplitudes, as before), $\tilde{\Delta}_{++}(k)$ and $\tilde{\Delta}_{--}(k)$. Energetic calculations show that the ground state is minimized with the states $\sim (k_x \pm i k_y)$~\cite{Anderson,Nishida}. Then we obtain

\begin{eqnarray}
\tilde{\Delta}_{++}(k) \sim (k_x + i k_y) \tilde{\Delta}_{++},
\\
\nonumber
\tilde{\Delta}_{- -}(k) \sim (k_x - i k_y) \tilde{\Delta}_{- -},
\end{eqnarray}
where $\tilde{\Delta}_{++}$ is given by Eq.~(\ref{pairamp}). The fact that we have calculated the p-wave pairing gaps induced by density fluctuations in the helicity basis allows us to write the {\it total} triplet p-wave pairing gaps in a unique expression,

\begin{eqnarray}
\label{total1}
\Delta_{++}^{total}(k) = \Delta_{++}(k) + \tilde{\Delta}_{++}(k),
\\
\nonumber
= \frac{(k_x + i k_y)}{k_{\perp}} \left(\Delta_p(k) + \tilde{\Delta}_{++} \right),
\end{eqnarray}
where $\Delta_p(k)$ and $\tilde{\Delta}_{++}$ are given by Eqs.~(\ref{gaps}) and (\ref{gapmais}), respectively, and

\begin{eqnarray}
\label{total2}
\Delta_{- -}^{total}(k)=  \Delta_{- -}(k)+ \tilde{\Delta}_{- -}(k),
\\
\nonumber
= \frac{(k_x - i k_y)}{k_{\perp}} \left(\Delta_p(k) + \tilde{\Delta}_{- -} \right),
\end{eqnarray}
where $\tilde{\Delta}_{--}$ is given by Eq.~(\ref{gapmenos}). 

Now we discuss the limiting cases of the dominant gaps (interactions) namely, the induced by RTSOC or the induced by density fluctuations ones. In the limit of very small spin-orbit coupling ($\alpha \simeq 0$), no matter the value of the s-wave interaction $g$, the second-order p-wave induced by density fluctuations will dominate. However, for small $g$ and strong spin-orbit coupling i.e., $\alpha \gg 1$ (or $\lambda \gg v_F$), the RTSOC-induced p-wave pairing gaps dominate.

%%%%%%%%%%%%%%%%%%%%%%%%%%%%%%%%%%%%%%%%%%%

\section{Conclusions}
\label{Conc}

In summary, we have calculated the triplet pairing gap in an imbalanced Fermi gas induced by both a RTSOC, and that from a second-order effect on the particle interaction $g$, due to density fluctuations. We have shown that the first-order contribution dominates in the limit of strong RTSOC. However, the latter prevails in the limit of zero RTSOC, in which case the former induced p-wave pairing gap (which comes from the first-order ``direct'' interaction) vanishes. 

We expect that with the current experimental techniques it will be possible soon to detect (intra-species) p-wave gaps in highly imbalanced systems. Then, the detection of the enhancement of the p-wave gaps could be done turning on a RTSOC and comparing the resulting images with the ones without the RTSOC synthetic gauge field. Since a ``weakly-paired'' p-wave superfluid with $p_x+ip_y$ symmetry in two dimensions is of special relevance because its vortices support zero-energy Majorana fermions and exhibit non-Abelian statistics~\cite{Read}, we hope our work can stimulate new investigations in this promising subject using our results.

%%%%%%%%%%%%%%%%%%%%%%%%%%%%%%%%%%%%%%%%%%%%
\section{Acknowledgments}

H. C. acknowledges the kind hospitality of CBPF were this work was done. He thank Drs. Jason Alicea, Lin Dong, Lianyi He, Eduardo Miranda and Leo Radzihovsky for useful discussions. This work was partially supported by CAPES, CNPq, FAPERJ,  and FAPEMIG (Brazilian Agencies). 

%\end{acknowledgments}

%%%%%%%%%%%%%%%%%%%%%%%%%%%%%%%%%%%%%%%%%%%%

\end{document}